\documentstyle[multicol,pre,aps,epsf]{revtex}

\begin{document}   

\draft				     

\title{Slow Manifold Structure and the Emergence of Mixed-Mode Oscillations}

\author{Andrei Goryachev, Peter Strizhak$^{\dag}$, and Raymond Kapral}
\address{Chemical Physics Theory Group, Department of
Chemistry, \\ University of Toronto, Toronto, ON M5S 3H6, Canada}
\address{$^{\dag}$ L.V. Pisarzhevskii Institute of Physical Chemistry, \\
Ukrainian Academy of Sciences, p. Nauki 31, Kiev, Ukraine, 25203}

\maketitle
\begin{abstract}		  

A  detailed  study  of the slow manifold of a model exhibiting
mixed-mode  oscillations  is  presented.  A  scenario  for the
emergence  of  mixed-mode  states which does not involve phase
locking   on   a   2-torus   is   constructed.  We  show  that
mixed-modes  correspond  to  the  periodic  orbits embedded in
the   horseshoe-type   strange   set   and   demonstrate   how
transformations  of  this  set  determine  the  transitions of
mixed-mode states into each other.

\end{abstract}

\pacs{ }

\begin{multicols}{2} 

\narrowtext 

\section{Introduction}\label{intro}

Oscillatory  behavior  exhibited by nonlinear chemical systems
often  takes  the  form  of  periodic  mixed-mode oscillations
consisting  of  $L$  large  amplitude oscillations followed by
$S$  small  amplitude oscillations. The distinctive difference
between  large  and  small oscillations allows one to classify
the  mixed-mode  oscillatory  states  by integer pairs $(L,S)$
and  assign  them  the symbol $L^{S}$. Ordered progressions of
such  periodic  patterns separated by chaotic oscillations are
observed  as  control  parameters are varied. The progressions
of   different   mixed-mode   forms  are  known  to  be  Farey
sequences described by Farey arithmetic \cite{ma-sw}.

Since  their  discovery  in the Belousov--Zhabotinsky reaction
\cite{huds},  mixed-mode  oscillations  have  been  found in a
broad  range  of  chemical  and biological \cite{gen} systems.
Their  ubiquity  suggests  the possibility of the existence of
a   common  dynamical  systems  theory  origin.  A  number  of
authors  \cite{ma-sw,roux}  identified experimentally observed
mixed-mode  oscillations  with  periodic  motions arising as a
result  of  phase-locking  on  a  2-torus. This hypothesis was
based  mainly  on  the  fact  that  the  periodic states on an
invariant  torus  form  Farey sequences as a control parameter
runs   through   the   quasiperiodic   route   from   period-1
oscillations  to  chaos.  Numerical evidence for the existence
of  quasiperiodicity  and torus oscillations was obtained from
studies    of   model   systems   which   exhibit   mixed-mode
oscillations \cite{bark1,lart}.

However,   there   are  results  which  show  that  mixed-mode
oscillations   cannot  be  attributed  to  the  phenomenon  of
quasiperiodicity    in    all   systems.   In   a   study   of
electrodissolution   of   copper   Albahadily   {\it  et  al.}
\cite{albah}  demonstrated  that  the mixed-mode states emerge
in  parameter  regions  different  from those where a torus is
stable.   The   formation   of   mixed-mode  oscillations  was
observed  at  a  point  well  beyond  the  range of parameters
between   the   secondary   Hopf  bifurcation  and  the  torus
break-up  bifurcation,  where the mixed-mode states should lie
according  to  the hypothesis of quasiperiodicity. In a series
of  electrochemical  studies Koper {\it et al.} \cite{kop-gas}
found  that  the  mixed-mode  oscillations  in Farey sequences
observed  in  their  study  are  separated  by  chaotic states
which  resemble  random  mixtures  of  the  adjacent  periodic
patterns  rather  than  by  quasiperiodic oscillations densely
covering  the  surface of a 2-torus. In a numerical study of a
model  system  Koper  \cite{koper}  found  a  parameter domain
where  both  quasiperiodic oscillations lying on a 2-torus and
a  large-amplitude  mixed-mode  oscillatory  state  which does
not  belong  to  the torus are stable simultaneously. In order
to   interpret  the  results  of  the  experimental  study  in
Ref.~\cite{albah}    Ringland   {\it   et   al.}   \cite{ring}
constructed   a  family  of  two-extremum  Z-maps  capable  of
producing  Farey  sequences  in  the  limiting  case  when the
central  segment  of  the  Z-map  is vertical. They have shown
that  one  may obtain Farey sequences which are not related to
phase locking on a torus.

In  this  paper  we  present  results  of  investigations of a
model  proposed  earlier  for  the  qualitative description of
the     mixed-mode     oscillations     observed     in    the
Belousov--Zhabotinsky    reaction   \cite{strik}.   From   the
analysis  of  the  transformation  of the slow manifold of the
model  as  system  parameters  are  changed we demonstrate the
gradual  emergence  of  a  horseshoe-type strange set in which
all  the  mixed-mode  states  are embedded as periodic orbits.
On    the    basis   of   this   analysis   we   construct   a
three-dimensional   phase   space   picture   describing   the
emergence   and   bifurcations   of  mixed-mode  states.  This
scenario  does  not  involve  a  torus  and  appears  to be an
alternative to the well-established quasiperiodic scenario.

In  Sec.~\ref{model}  we  introduce  the  model  and present a
brief   discussion   of   its   attractors   and   bifurcation
structure.  Section~\ref{shoe}  is  devoted  to  the  detailed
analysis  of  the  slow  manifold  and  a  description  of its
important  properties.  We  introduce  a technique that allows
us  to  construct  the  slow  manifold  and demonstrate how it
turns  into  a  horseshoe  by  stretching  and  folding  . The
repulsive  flow  that defines a particular shape of mixed-mode
oscillations  is  introduced  and discussed in some detail. In
Sec.~\ref{mmo}  we  show that when the slow manifold is folded
into  a  horseshoe  interacting  with  the repulsive flow, the
mixed-mode  oscillations  arise  naturally  at the point where
the    large-amplitude,   period-1   oscillation   loses   its
stability  through  a saddle-node bifurcation. The relation of
the  phase  flow  on  the horseshoe to the Z-map introduced by
Ringland   {\it   et   al.}   \cite{ring}   is  discussed.  In
Sec.~\ref{conc}   we   conclude  that  the  existence  of  the
repulsive  flow  allows  one  to  consider this horseshoe as a
representative  of  a  distinct subclass of strange sets whose
periodic windows are organised according to Farey sequences.

\section{Description of the model}  \label{model}

\begin{figure}[htbp]
\begin{center}
\leavevmode
\epsffile{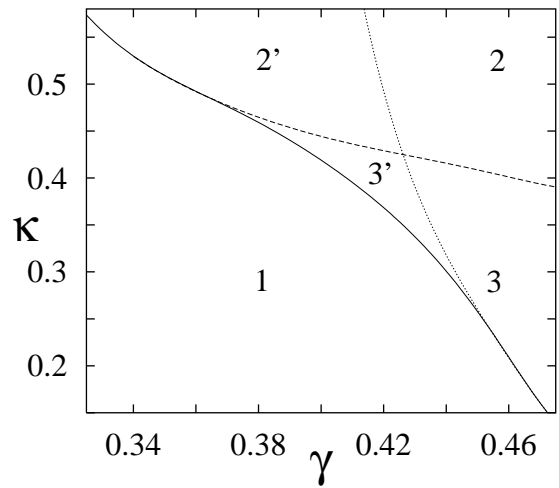}
\end{center}
\caption{Bifurcation   phase   diagram:  1  ---  steady  state
(focus),  2  ---  period-1,  3  ---  mixed-mode  oscillations.
Separating  lines  correspond  to the Hopf bifurcation (dots),
transition   ${\bf   1^0}   \leftrightarrow   {\bf  \infty^1}$
(dashes),  and  crash  onto  the  focus  (solid).  Both steady
state  and  oscillations  are  stable  in  subdomains $2'$ and
$3'$.}
\label{fig01}
\end{figure}

We  consider  a  three-variable  extension \cite{foot1} of the
Boissonade--De~Kepper model \cite{boiss},
\begin{eqnarray}
\label{sys}
\dot{x} & = & \gamma (y-x^3+3\mu x)\, , \nonumber \\
\dot{y} & = & -2\mu x -y-z+\beta \, ,\\
\dot{z} & = & \kappa (x-z)\, .	    \nonumber
\end{eqnarray}				     
System  (\ref{sys})  was  originally  proposed in \cite{strik}
in  order  to  obtain  a  qualitative description of transient
mixed-mode    oscillations    in   the   Belousov--Zhabotinsky
reaction.  Depending  on  the  values  of  $\mu$  and $\beta$,
(\ref{sys})  possesses  either one or three steady states. The
change in the number of stationary points occurs on the curve
\begin{equation}			       
\label{det}
\left(\frac{\beta}{2}\right)^2 = \left(\frac{\mu - 1}{3}\right)^3\, ,
\end{equation}
where  two  steady  states coalesce or emerge in a saddle-node
bifurcation.  At  $\mu  =1,\, \beta =0$, where two branches of
the  curve  (\ref{det})  meet  in a cusp point, the pitch-fork
bifurcation  takes  place.  Note  that the parameters $\gamma$
and  $\kappa$  only  rescale  the  corresponding components of
the  vector  field  and  do not influence the positions of the
nullclines   and   steady   states  in  phase  space.  In  the
following   discussion  the  geometric  parameters  $\mu$  and
$\beta$  are  fixed  $(\mu =2.0,\,\beta =-0.4)$ while $\gamma$
and  $\kappa$  are  varied.  With  this  choice  of  $\mu$ and
$\beta$  system  (\ref{sys}) has only one steady state at $x_f
=  z_f  =  -1.159705,\, y_f=5.398524$, which is a stable focus
in  a  domain  of  the parameter plane $(\gamma,\kappa)$ lying
to    the   left   of   the   Hopf   bifurcation   line   (see
Fig.~\ref{fig01}),  and  a saddle-focus otherwise. As the real
characteristic  exponent  of  the  steady  state  is  negative
everywhere  on  $(\gamma,\kappa)$, no second order bifurcation
is possible at the chosen values of $\mu$ and $\beta$.

Figure~\ref{fig01}     shows     the    partition    of    the
$(\gamma,\kappa)$  plane  into  domains  of different types of
dynamical  behavior  according  to  the  results  of numerical
integration.   In   domain  1  a  stable  focus  is  the  only
attractor   of   the   system   (\ref{sys}).  Large  amplitude
period-1    $({\bf   1^0})$   oscillations   with   a   strong
relaxational  character  at large $\gamma$ are found in domain
2.
Finally,  mixed-mode  oscillations  are  observed  in  the
wedge-shaped   domain   3.   Transitions   from   period-1  to
mixed-modes  take  place on an almost flat curve (shown by the
dashed  line  in  Fig.~\ref{fig01})  where  the  trajectory of
system  (\ref{sys})  performs a small-amplitude excursion only
once  on  the  entire  time interval $t\in (-\infty,+\infty)$.
This  transition  can  be  represented  symbolically  as ${\bf
1^0}  \leftrightarrow  {\bf \infty^1}$. Continuing to traverse
domain  3  by  decreasing  $\gamma$  at  constant $\kappa$ one
observes   pruned   Farey  sequences  of  periodic  mixed-mode
oscillations  ${\bf  \infty^1}\rightarrow {\bf n^1}\rightarrow
{\bf  1^1}\rightarrow  {\bf  1^n}\rightarrow  {\bf 1^\infty} $
interspersed  by  narrow, yet detectable, spans of chaos. This
periodic-chaotic  progression  ends  suddenly  in a crash onto
the  focus  at the point where the line separating zones 1 and
3  is  crossed. At large $\kappa$ ($\kappa \geq 0.54 $), where
the  crash  bifurcation  curve appears to be tangental to that
for   the   ${\bf   1^0}   \leftrightarrow   {\bf   \infty^1}$
transition,   both  bifurcations  occur  within  an  extremely
narrow     parameter    span    and    one    observes    hard
termination/emergence      of     large-amplitude     periodic
oscillations.   Another  limiting  case  is  observed  at  low
$\kappa$  ($\kappa  \leq  0.20 $) where the crash curve merges
with  the  Hopf  bifurcation  line. Here the crash occurs from
pattern   ${\bf   1^n},\,   n\gg  1$  which  forms  an  almost
homoclinic   connection  to  the  focus  undergoing  the  Hopf
bifurcation.

\begin{figure}[htbp]
\begin{center}
\leavevmode
\epsffile{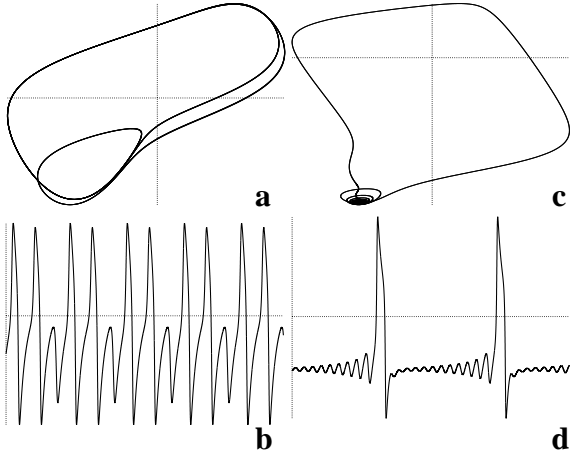}
\end{center}
\caption{Projections  of  the phase portraits on $(x,z)$ plane
and   corresponding   $x(t)$  time  series  for  two  periodic
mixed-mode  states:  (a,b)  ---  ${\bf  2^1}$ for $\kappa=0.5,
\gamma  =  0.35516$; (c,d) --- ${\bf 1^{12}}$ for $\kappa=0.2,
\gamma=0.471$.}
\label{fig02}
\end{figure}

As  $\kappa$  varies  the  appearance of the phase portrait of
the  mixed-mode  states  changes considerably. At large values
of  $\kappa$  both  small-  and large-amplitude excursions lie
almost   on   the   same   plane.   This  type  of  mixed-mode
oscillation,  referred  to  type-1 according to classification
given  in  \cite{kop-gas},  is shown in Fig.~\ref{fig02}(a,b).
With   a   decrease   in  $\kappa$  the  reinjection  part  of
trajectory  falls  closer  to  the focus and the angle between
the  planes  of small- and large-amplitude oscillations grows.
At  $\kappa\leq  0.2$  the  phase  portrait of the oscillation
acquires   all   of  the  features  characteristic  of  type-2
mixed-mode   states.   Here   the   large   loops  are  mainly
orthogonal  to  the  small  ones  and  the  reinjection occurs
along  the  one-dimensional  stable manifold of the focus (see
Fig.~\ref{fig02}(c,d)).

\section{Construction of the phase flow} \label{shoe}

The  parameter  region  in  which complex periodic and chaotic
oscillations  are  found  can  be  partitioned into subdomains
corresponding  to  particular periodic mixed-mode states $L^S$
and  their  chaotic  mixtures.  We  shall study the mechanisms
responsible  for  the  bifurcations  of  one  mixed-mode state
into  another  by  analysing transformations of the phase flow
on the system's slow manifold.

The  concept  of a slow manifold relies on the assumption that
the  relaxation  of  an  arbitrary  initial  condition  to the
attractor  has  two characteristic time scales: the trajectory
quickly  reaches  certain  subset  of the phase space and then
slowly  approaches  the  attractor  within this subset. In the
case  of  a  strongly  dissipative system with three dynamical
variables,  the  slow  set  is  usually  represented  by  a 2D
manifold  which  is  often  folded  in  the 3D phase space. To
avoid  complications  arising from nontrivial embedding of the
slow  manifold,  we  choose for its construction the region of
large  $\kappa$  where  the  slow  manifold  appears to have a
relatively simple organization.

\subsection{Slow manifold}  \label{shoe1}

\begin{figure}[htbp]
\begin{center}
\leavevmode
\epsffile{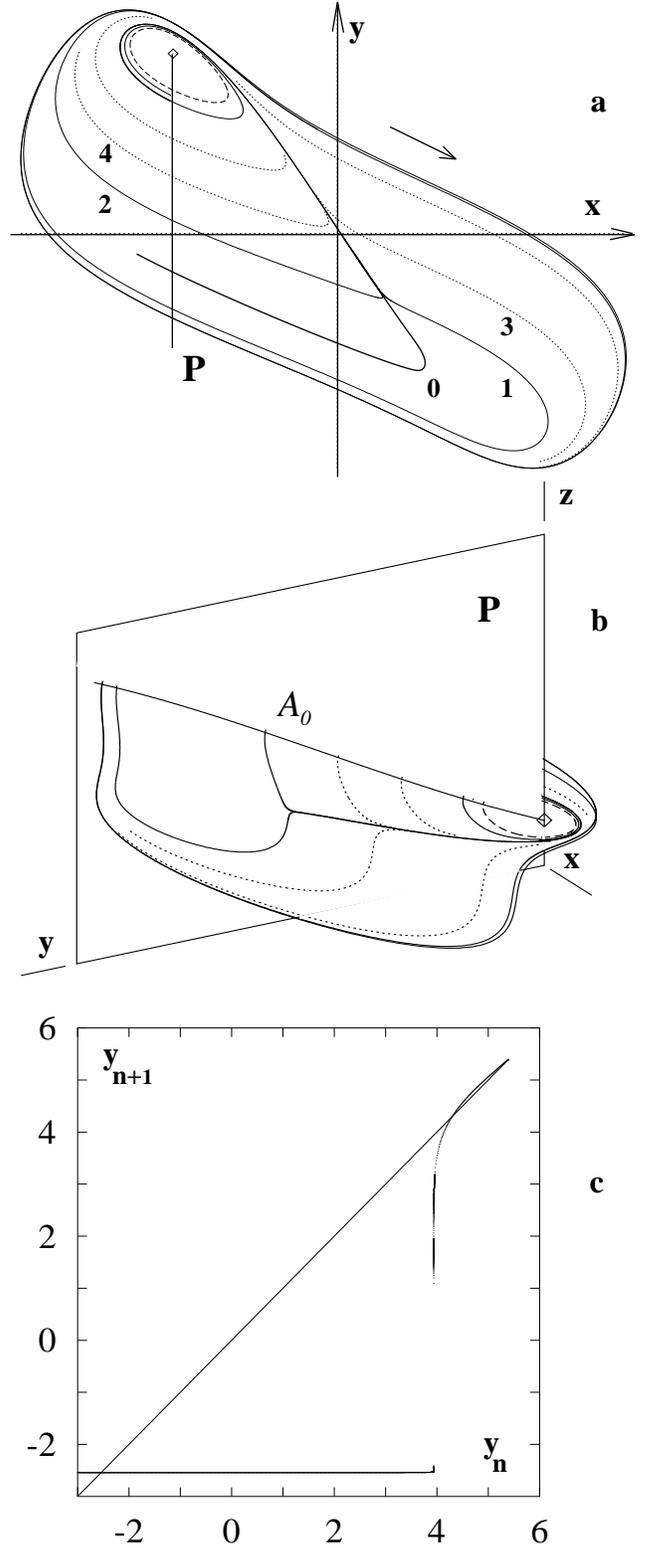}
\end{center}
\caption{Phase  flow  on  the slow manifold. (a) Projection of
the  flow  on  $(x,y)$ plane. (b) Mutual positions of the slow
manifold  and  the  surface-of-section $P$ in phase space. (c)
First return map for the flow initiated on $A_0$.}
\label{fig03}
\end{figure}

Figure~\ref{fig03}(a)  presents  the  projection  of the phase
flow  on  the  $(x,y)$  plane  for  $\kappa=0.55$ and $\gamma=
0.39$.  For  these  parameter  values  deep within domain $2'$
the  system  (\ref{sys})  is  bistable.  Apart from the stable
focus,  two  stationary  periodic  solutions  exist: a stable,
large-amplitude,  period-1
limit  cycle  (solid  line)  and a
saddle-type,   small-amplitude   cycle   (dashed   line). The
planar,  concentric  organization  of  these cycles around the
focus  suggests  that  the  slow  manifold  containing them is
flat.   Trajectories   that   leave   the   saddle   cycle  at
$t\rightarrow  -\infty$  and approach stable cycle or focus at
$t\rightarrow   +\infty$  lie  in  the  slow  manifold.  These
trajectories  are  easy  to calculate and one can use them for
the  construction  of  a  Poincar\'{e}  section  of  the  slow
manifold.   We   choose   $P=\{x=x_f,\,   y<y_f\}$   shown  in
Fig~\ref{fig03}(a,b)  as  a  surface of section. Intersections
of  the  transient trajectories with $P$ yield a set of points
which,  after  interpolation  by  a smooth function $z = z(y)$
\cite{foot2},  becomes  a  continuous zero-order approximation
$  A_0=\{x=x_f,\,  y<y_f,\,  z  =  z(y)\}$ to the Poincar\'{e}
section of the slow manifold.

Figure~\ref{fig03}(b)  shows  a  3D phase portrait of the same
trajectories  as  in  Fig.~\ref{fig03}(a),  as well as $P$ and
$A_0$.  The  intersections of the flow initiated on $A_0$ with
$P$   can  be  used  both  to  improve  the  accuracy  of  the
approximation  and  to  construct  the  first  return map (see
Fig.~\ref{fig03}(c))   which   quantitatively   describes  the
phase flow on the slow manifold.
								  
A   complete  description  of  the  phase  flow  on  the  slow
manifold  can  be  obtained  from  the comparison of the phase
portrait  and  the  first  return  map.  Consider  the example
presented  in  Fig.~\ref{fig03}(a,c).  The  first  return  map
$y_{n+1}=f(y_n)$  intersects  the  bisectrix  in  three points
which  correspond  to  the stable limit cycle $(y_c=-2.5437)$,
the   saddle   cycle  $(y_s=4.2998)$,  and  the  stable  focus
$(y_f=5.3985)$.  The  nearly  rectangular  shape  of  the  map
implies  that  the  slow  manifold  comprises  areas of strong
expansion  (the  neighborhood  of  $y_0=3.9370$, where the map
is  vertical)  and  contraction  (horizontal part of the map).
In  the  course of its evolution a trajectory initiated in the
vicinity   of   the   saddle   cycle   executes  a  number  of
small-amplitude  loops  around  the  saddle  cycle and then in
one  turn  reaches the neighborhood of the stable limit cycle.
Whether   the  trajectory  approaches  the  limit  cycle  like
trajectory   3   performing   small
\begin{figure}[htbp]
\begin{center}
\leavevmode
\epsffile{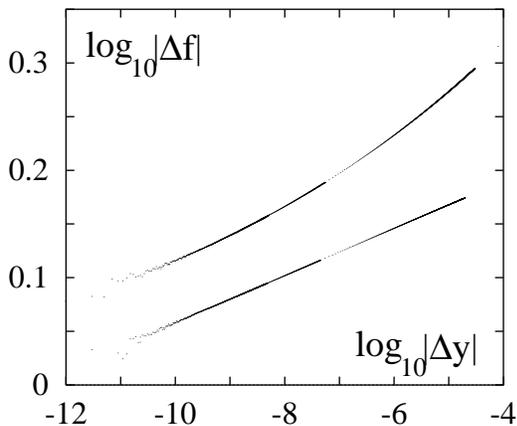}
\end{center}
\caption{Scaling  of  $f(y)$  at the point $y_0$ where tangent
is vertical.}
\label{fig04}
\end{figure}
transient  loop  or  like
trajectory   4  which  makes  a  large  loop  depends  on  its
position relative  to  a  certain boundary trajectory 0 which
can  be  identified  with  the point $y_0$ on the first return
map  where  $f(y)$  has a vertical tangent. Analysis of $f(y)$
in  a  small neighborhood of $y_0$ shows that the trajectory 0
is  superunstable;  {\it  i.e.},  the neighboring trajectories
deviate     from     it     faster     than     geometrically.
Figure~\ref{fig04}   shows   $\log_{10}   |\Delta  f|$  versus
$\log_{10}  |\Delta  y|$,  where  $\Delta  f=f(y)-f(y_0)$  and
$\Delta   y=y-y_0$,  for  both  left  $y<y_0$  (lower  set  of
points)  and  right  $y>y_0$  (upper  set) branches of the map
shown  in  Fig.~\ref{fig03}(c).  One  can see that for $\Delta
y$  sufficiently  small, $\Delta f(\Delta y)$ is characterized
by  the  power  law  $\Delta  f  \propto  (\Delta y)^{\alpha}$
where  $\alpha  \approx  0.022 \ll 1$. This implies an extreme
divergence  of  trajectories  which  is  also reflected in the
phase  portrait.  The  initial  conditions  $(x_f,y_1,z(y_1)),
\;(x_f,y_2,z(y_2))$  on  $P$  for  trajectories 1 and 2 (solid
lines     in    Fig.~\ref{fig03}(a,b))    were    such    that
$y_1=y_0-\delta    /2,\,y_2=y_0+\delta   /2$   where   $\delta
=10^{-12}$.   As   Fig.~\ref{fig03}   shows,   the  separation
between   these  two  trajectories  on  their  return  to  the
surface   of   section   grows   by  $\approx  12$  orders  of
magnitude.  In  fact,  trajectories 3 and 4 constitute another
pair with $\delta =10^{-10}$.

The  existence  of the superunstable trajectory results in the
observed  dichotomy  of  trajectories  belonging  to  the slow
manifold.  A  2D manifold intersecting the slow manifold along
trajectory  0  plays  similar  separating role in the 3D phase
space.  Although  the  global structure of this repulsive flow
in  the  phase  space  is  very  complicated,  the flow can be
easily  determined  locally  by the property that it separates
trajectories  making  large  and small loops while approaching
the attractor.

\subsection{Birth of horseshoe} \label{shoe2}

\begin{figure}[htbp]
\begin{center}
\leavevmode
\epsffile{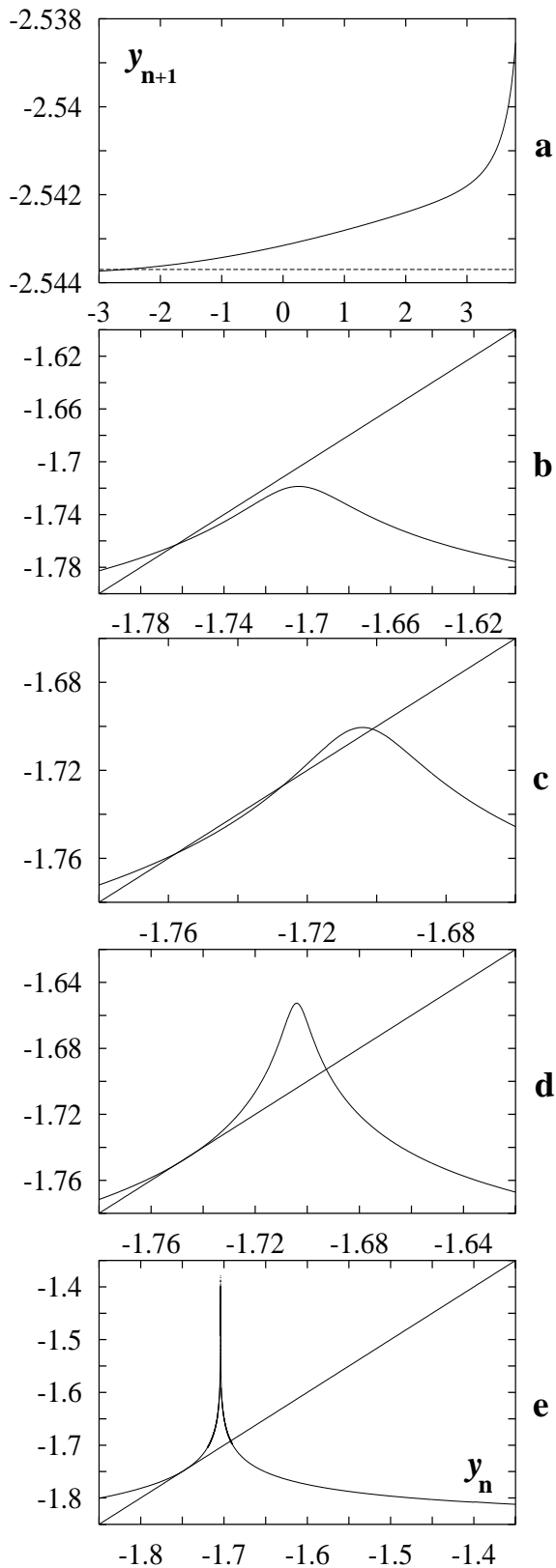}
\end{center}
\caption{Transformation   of   the   first   return   map   at
$\kappa=0.55$  and  decreasing  $\gamma$:  (a)  $\gamma=0.39$,
(b)    $\gamma=0.332241$,    (c)    $\gamma=0.3322399$,    (d)
$\gamma=0.3322392$, (e) $\gamma=0.332239127$.}
\label{fig05}
\end{figure}

Figure~\ref{fig05}(a)  shows  an  expanded  view  of  the flat
part  of  the  first return map $f(y)$ in Fig.~\ref{fig03}(c).
The  slope  of  the  map is everywhere positive and, thus, the
map   is   invertible.   As  $\gamma$  decreases  at  constant
$\kappa=0.55$,  a  smooth  transformation of the slow manifold
takes  place,  resulting  in  the loss of invertability of the
map.  A  fragment  of  map  constructed  for $\gamma=0.332241$
(see  Fig.~\ref{fig05}(b))  demonstrates  that  $f(y)$  is not
monotonic  and  a  smooth maximum arises near the stable limit
cycle.  At  $\gamma=0.33224025$  the  map  is  tangent  to the
bisectrix  and  a new pair of stable and unstable limit cycles
emerges  on  the  slow  manifold (cf. Fig.~\ref{fig05}(c)). As
the  map's  maximum  grows  it  sharpens and the newborn limit
cycle    looses   its   stability   through   a   cascade   of
period-doubling   bifurcations.   At  $\gamma=0.3322392$  (see
Fig.~\ref{fig05}(d))  a  chaotic  attractor  coexists with the
period-1  limit  cycle  and focus. Although the entire cascade
to  chaos  occupies  very  narrow  span of $\gamma$, two first
period-doubling  bifurcations  were resolved. The tristability
of  the  system breaks down at a certain value of $\gamma$ and
every  trajectory  initiated  near  the  maximum of the map is
attracted   to  the  period-1  limit  cycle  after  a  chaotic
transient of varying length.
Figure~\ref{fig05}(e)  shows  a  part  of the first return map
for  $\gamma_c=0.332239127$  where oscillations suddenly crash
on  the  focus.  Two  important characteristic features of the
flow  can  be inferred from the map. First, the maximum of the
map  acquires  a  cuspoid shape signalling of the formation of
a  highly  compressed  fold on the slow manifold. Second, from
the  tangency  of  the map and bisectrix, one can see that the
stable  period-1  limit  cycle  and the large-amplitude saddle
cycle  are  near the point of coalescence. The annihilation of
this  pair  in  a  saddle-node  bifurcation  gives  rise to an
intermittent  chaotic  attractor which has the appearance of a
thick  limit  cycle.  A  decrease  in  the  control  parameter
$\gamma$  leads  to the crash of chaotic oscillations onto the
focus.    Summarising    the    information    presented    in
Fig.~\ref{fig05}(a-e),  one  can  conclude  that  as  $\gamma$
changes  through  the  interval  $[0.39,\gamma_c]$,  the  slow
manifold  transforms  from  a  planar  surface  into  a folded
fractal   structure   with  infinitely  many  leaves  (Smale's
horseshoe).

\begin{figure}[htbp]
\begin{center}
\leavevmode
\epsffile{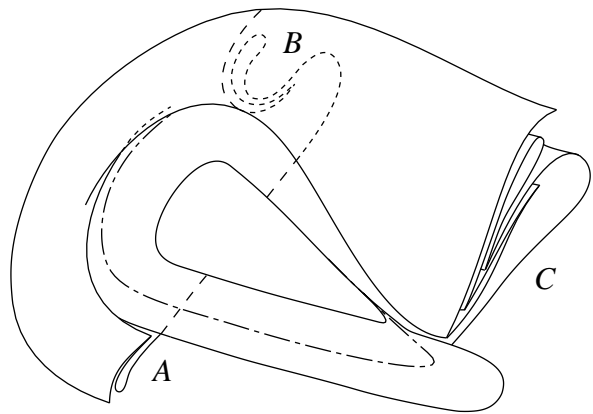}
\end{center}
\caption{Geometry   of   the   folded   slow   manifold.   The
superunstable trajectory is shown by dot-and-dash line.}
\label{fig06}
\end{figure}

Making  use  of  several auxiliary Poincar\'{e} sections it is
possible  to  construct  a  3D  picture  of the slow manifold.
Figure~\ref{fig06}  schematically  represents the manifold and
three  of  its  Poincar\'{e}  sections  at parameters close to
the  crash  bifurcation. Part of the manifold situated between
sections  $C$  and  $A$ is removed to facilitate visualization
of  its  geometry.  In  section  $A$  one sees a triply-folded
curve  which  corresponds to the first structural level of the
manifold.  Note  the  difference between the two folds of this
curve:  while  the  outer  fold  has a smooth parabolic shape,
the  inner  fold  is  a  sharp cusp. Increasing the resolution
considerably  (by  $\approx$  5 orders of magnitude) one would
be  able  to  find  that  the  curve  is,  in  fact, a densely
compressed  fold  constituted  by  seven  leaves of the second
structural  level,  and  so  on.  To describe one iteration of
the  manifold  stretching  and  folding  we  do  not  need  to
consider   these   higher  structural  levels.  Following  the
evolution  of  points in section $A$ under the flow (clockwise
in   Fig.~\ref{fig06})   one  encounters  a  pleat  where  the
manifold  as  a  whole  begins to fold into an $S$-shaped form
(section  $B$)  with  two approximately equal parabolic folds.
Prominent   stretching  of  the  flow  in  the  plane  of  the
manifold,
accompanied   by   strong   contraction   in   the
orthogonal  direction,  transforms  the manifold into the thin
roll  composed  of  seven  leaves  seen  in  section $C$. As a
result  of  the extreme inhomogeneity of compression the inner
fold   acquires  sharp  cuspoid  shape  while  the  outer  one
remains  smooth  and parabolic. As the flow returns to section
$A$  more  contraction along the vertical direction occurs and
the  second  order  structure  shown  in  section  $C$  is not
discernible.

\section{Mixed-Mode oscillations}  \label{mmo}

In  Sec.~\ref{shoe}  we described two important characteristic
features  of  the  phase  flow of system (\ref{sys}). Firstly,
at  the  parameter  values  in  the neighborhood of mixed-mode
oscillations,   the   slow   manifold   is   folded   into   a
horseshoe-type  set  with  a  fractal, multi-leaved structure.
Secondly,   the   repulsive   flow   divides   the   transient
trajectories  into  two  distinct classes. Intersection of the
repulsive  flow  and  the slow manifold yields a superunstable
trajectory   with   peculiar   dynamical  properties.  In  the
present   section   we   show   how  mixed-mode  states  which
correspond  to  periodic  orbits  embedded  in  the  horseshoe
emerge and bifurcate into each other.

\subsection{Emergence of mixed-mode oscillations} \label{mmo1}

Consider  a  path  in  the  $(\gamma,\kappa)$  parameter plane
which  starts  in  domain  2 or $2'$ and ends in the region of
mixed-mode  oscillations  3  (see  Fig.~\ref{fig01}).  As  the
domain  of  large-amplitude period-1 oscillations is traversed
towards    the    curve    corresponding    to    the    ${\bf
1^0}\leftrightarrow{\bf     \infty^1}$     transition,     the
modifications    of    the    slow   manifold   described   in
Sec.~\ref{shoe2}  take  place.  Before the first mixed-mode is
formed  an  important transformation of the manifold occurs at
the  point  where  a cusp-shaped fold developes. Inspection of
Fig.~\ref{fig06}   shows  that  the  manifold  cannot  support
periodic  oscillations  with  small-amplitude  loops since its
folded  part  lies  on  the  ``large-amplitude''  side  of the
repulsive  flow  (only  the  superunstable trajectory is shown
in  Fig.~\ref{fig06}).  In  order  for orbits with both small-
and  large-amplitude  excursions  to exist, the repulsive flow
has to intersect the folded part of the horseshoe.

\begin{figure}[htbp]
\begin{center}
\leavevmode
\epsffile{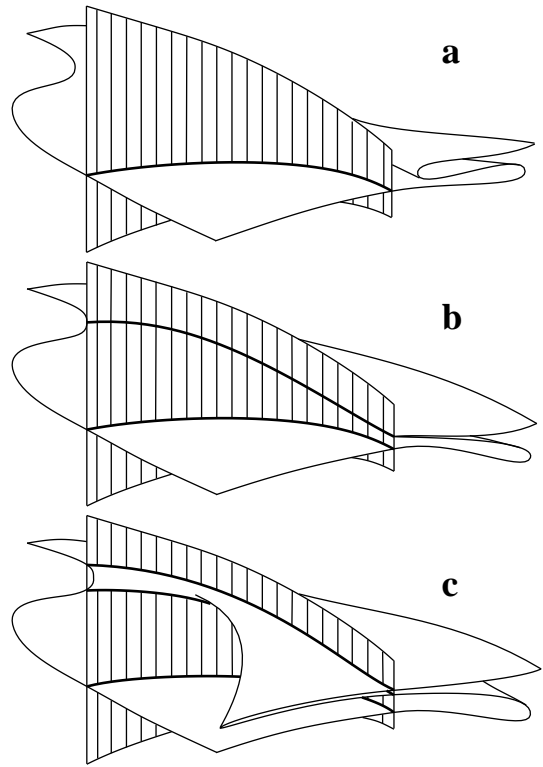}
\end{center}
\caption{Propagation   of   the   slow  manifold  through  the
surface   of   the   repulsive  flow  (hatched):  (a)  $\gamma
>\gamma_c$      ---      transversal     intersection;     (b)
$\gamma=\gamma_c$  ---  formation  of  a tangency; (c) $\gamma
<\gamma_c$ --- fold penetrates the surface.}
\label{fig07}
\end{figure}

Figure~\ref{fig07}  shows  segments  of  the manifolds cut out
by  two  orthogonal  planes  for  three  consequtive values of
$\gamma$  and  fixed  $\kappa$.  The  case  $\gamma >\gamma_c$
(Fig.~\ref{fig07}(a))  corresponds  to the generic transversal
position  of  the  manifolds  in which the growing fold of the
horseshoe  lies  on  one side of the repulsive flow. This case
is  described  by  a  first  return  map  of the type shown in
Fig.~\ref{fig05}(d).  At  $\gamma=  \gamma_c$,  in addition to
the  transversal  intersection,  a  tangency  between  the two
manifolds   occurs.  The  tangency  manifests  itself  by  the
formation  of  a  sharp  cusp  on  the  first  return map (see
Fig.~\ref{fig05}(e)).  For  $\gamma  <  \gamma_c$  the fold of
the  horseshoe  penetrates  the  surface  corresponding to the
repulsive  flow    
and drastically increases in size due to the
strong  divergence  from  the repulsive flow. One also sees in
Fig.~\ref{fig07}(c)  that  as  a result of this transformation
two  new  superunstable orbits (thick solid lines) are born on
the  slow  manifold.  Taking  into  the  consideration fractal
nature  of  the  horseshoe one could imagine an infinite stack
of  superunstable  orbits  born  in pairs when nested folds of
the  horseshoe  penetrate  the  surface  of the repulsive flow
one by one.

\begin{figure}[htbp]
\begin{center}
\leavevmode
\epsffile{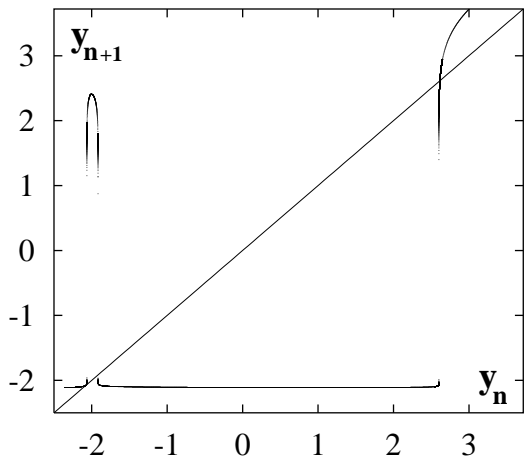}
\end{center}
\caption{First  return  map  at  the  point  of the ${\bf 1^0}
\leftrightarrow  {\bf  \infty^1}$  transition  ($\kappa=0.5,\;
\gamma=0.355403274$).}
\label{fig08}
\end{figure}

The  birth  of  new  superunstable  orbits is reflected in the
first  return  map as well. Figure~\ref{fig08} shows the first
return  map  constructed  with  $P$ as a surface of section as
described    in    Sec.~\ref{shoe1}   for   parameter   values
$\kappa=0.50,\,  \gamma_0=  .355403274  <  \gamma_c$  at which
the   transition   ${\bf   1^0}\leftrightarrow{\bf  \infty^1}$
occurs.  As  $\gamma_c$  is passed, the cusp on the map breaks
up  and  the  maximum  acquires normal parabolic shape. At the
same  time  the  height  of the maximum increases tremendously
and two segments with infinite slope appear on its sides.

The  analysis  of  the  map  shows  that  the transition ${\bf
1^0}\leftrightarrow   {\bf  \infty^1}$  occurs  simultaneously
with   annihilation   of   the   large-amplitude   stable  and
saddle-type  limit  cycles.  At  the point of coalescence they
produce  structurally  unstable  limit  cycle which is seen on
the  first  return  map as a point of tangency with bisectrix.
This  cycle  combines  properties  of both stable and unstable
limit  cycles.  Consider  a trajectory which departs from such
cycle  at  $t  \rightarrow -\infty$. In course of evolution it
inevitably   executes
one  small-amplitude  loop  around  the
focus.  Upon  the  completion  of  this loop the trajectory is
reinjected  into  the  stable  part of the neighborhood of the
cycle   which  is  eventually  approached  at  $t  \rightarrow
+\infty$.  Thus,  first  mixed-mode  state ${\bf \infty^1}$ is
characterized  by  the  orbit  homoclinic  to the structurally
unstable  cycle  and,  therefore, exists in a parameter domain
of  zero  measure.  For $\gamma < \gamma_0$ a finite number of
large-amplitude  loops  is executed before a trajectory visits
the  small  loop  and  successions  of  ${\bf n^1}$ mixed-mode
states with rapidly decreasing $n$ are oserved.

\subsection{Embedding of the mixed-mode orbits} \label{mmo2}

\begin{figure}[htbp]
\begin{center}
\leavevmode
\epsffile{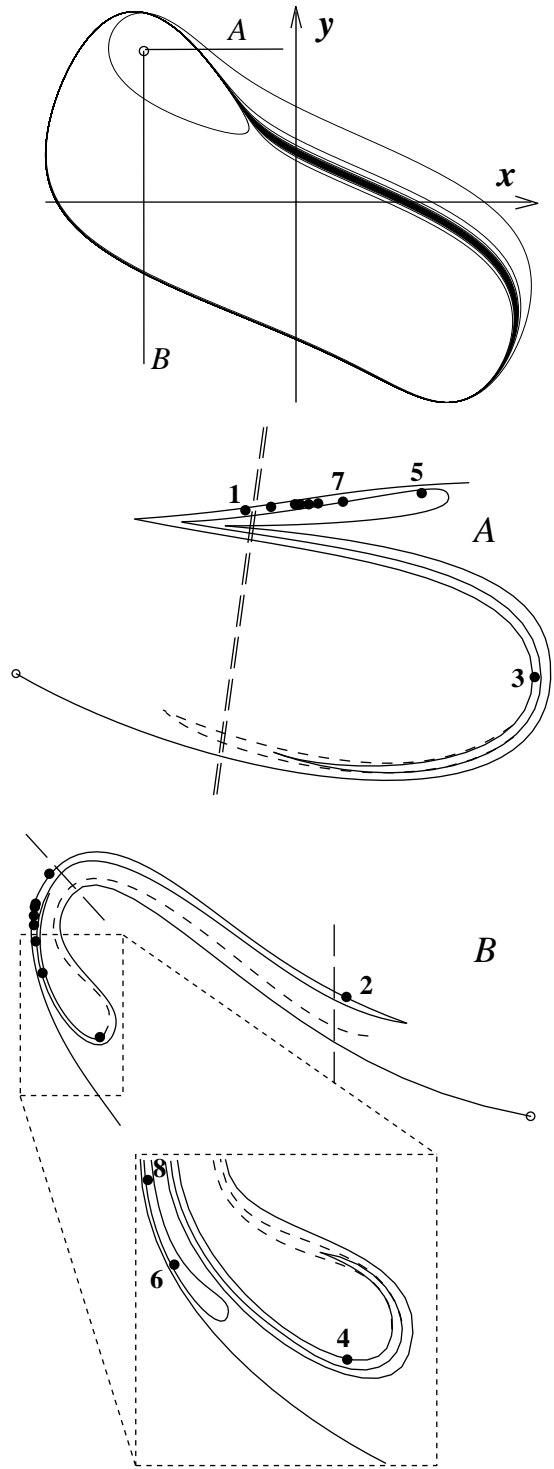}
\end{center}
\caption{Projection  of  ${\bf 46^1}$ orbit onto $(x,y)$ plane
and  its  embedding  in  the  slow manifold (see discussion in
the text).}
\label{fig09}
\end{figure}

It  is  possible  to  establish  how  a  particular mixed-mode
orbit     is    incorporated    in    the    slow    manifold.
Figure~\ref{fig09}  shows  the  $(x,y)$  projection of a ${\bf
46^1}$  orbit  and  its  embedding  in  the  horseshoe.  As in
Fig.~\ref{fig06},  the  Poincar\'{e}  sections of the manifold
are  shown  at  different  levels  of  resolution  of the fine
structure.   While   section   $A$   presents  the  first  two
structural  levels,  in  section $B$ only first level is shown
in  full.  Two  lines  of phase points parallel to the surface
of  the  repulsive  flow and separated from it by $\delta =\pm
0.5\times   10^{-12}$   are  selected  in  section  $A$  (long
dashes).  The  evolution  of these points shown in section $B$
demonstrates  the  extreme  divergence  of trajectories caused
by the presence of the repulsive flow.

Orbit   ${\bf  46^1}$  consists  of  a  dense,  flat  band  of
large-amplitude  loops  which consecutively shrink in size and
a  small  loop where the trajectory segment is reinjected into
the  large-amplitude  band.  Projecting  points  in  which the
orbit  intersects  planes $A$ and $B$ (shown by thick dots) on
the  corresponding  Poincar\'{e} sections of the horseshoe one
can  obtain  insight  into  the mechanism of this reinjection.
At  a  certain  time the trajectory appears to the left of the
repulsive  flow  surface  (point  1 in $A$). This causes it to
execute  a  small  loop  (point 2 on the cusp-shaped tongue in
$B$).  Next  return  of the trajectory into section $A$ occurs
in  point  3,  separated  far from 1. Following the subsequent
numbers    one   sees   how   the   orbit   returns   to   the
large-amplitude  band,  gradually ascending leaves of the slow
manifold.  The  same  mechanism  holds  for  all  ${\bf  n^1}$
states,   although  escape  on  the  small  loop  occurs  from
different leaves.

The   analysis  of  the  reinjection  scheme  shows  that  the
horseshoe  (solid  lines  in  Fig.~\ref{fig09}) cannot support
mixed-mode  oscillations  with  more  than one small-amplitude
loop.  Indeed,  regardless  of  how far to the left one shifts
the  position  of  point  1,  first  return  of  a  trajectory
initiated  at  1 into the surface of section (point 3) lies to
the  right  of  the  repulsive flow. This situation changes as
the  cusp-shaped  tongue, to which point 3 belongs, penetrates
the  surface  of the repulsive flow through the formation of a
tangency   as  described  in  Sec.~\ref{mmo1}.  (Corresponding
changes  on  both  sections are shown by medium-dashed lines.)
As  soon  as  a  similar  tongue  of a higher structural level
passes   through   the  surface  of  the  repulsive  flow,  an
opportunity  for  the  existence  of the third small-amplitude
loop  arises,  and  so  on.  Thus,  changing  position  of the
horseshoe  relative  to  the  repulsive  flow  one  can obtain
mixed-mode states $L^S$ with any desired $L$ and $S$.

\subsection{Relation to Z-map}

\begin{figure}[htbp]
\begin{center}
\leavevmode
\epsffile{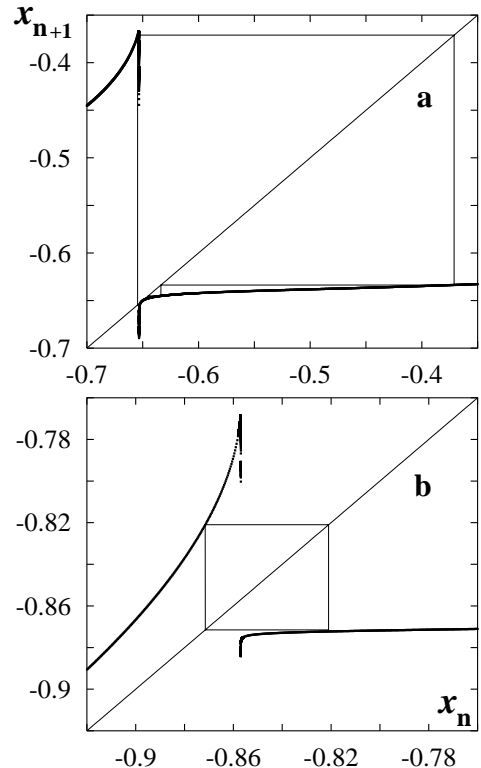}
\end{center}
\caption{First    return   maps   $f_Z(x)$   constructed   for
$\kappa=0.4$   with   $y=0.4$  as  a  surface-of-section:  (a)
$\gamma=0.46204$,  (b)  $\gamma=0.44$. Orbits ${\bf 46^1}$ and
${\bf 1^1}$ are shown with thin solid lines.}
\label{fig10}
\end{figure}

At   small   values   of  $\kappa$,  where  the  bulk  of  the
mixed-mode    domain    lies,   the   surface-of-section   $P$
introduced  in  Sec.~\ref{shoe1}  becomes  unsuitable  for the
construction  of  the  first  return  map  due to the strongly
bent  shape  of  the  corresponding  Poincar\'{e} section (see
section   $B$  in  Fig.~\ref{fig09}).  Instead,  one  can  use
sections    $y=$    const    similar   to   section   $A$   in
Fig.~\ref{fig09}.  Figure~\ref{fig10}  presents  two  examples
of  the  first  return  map  $x_{n+1}=f_Z(x_n)$ constructed in
the  approximation  based  on  the upper leaf of the horseshoe
for  parameter  values  corresponding  to ${\bf 46^1}$ (a) and
${\bf 1^1}$ (b) mixed-mode states.

As  one  can  see,  maps of this type share a number of common
features.  The  map  consists  of  two  branches with positive
slope  and  an  extremely  steep,  negatively-inclined segment
which  joins  the  branches. At the point corresponding to the
superunstable  orbit  the map has an infinite slope and can be
locally   described   by   an   exponent   as   discussed   in
Sec.~\ref{shoe1}.   Any   mixed-mode   state   $L^S$   may  be
represented  on  this  map  by  a periodic trajectory with $L$
iterates  on  the  right  branch  and $S$ iterates on the left
branch.  This  property  of  the  map  makes  it  particularly
suitable  for  the illustration of mixed-mode bifurcations. At
parameter   values   corresponding  to  the  transition  ${\bf
1^0}\leftrightarrow{\bf   \infty^1}$   the   right  branch  is
tangent  to  the  bisectrix. As an intermittent channel opens,
the  trajectory  funnels  through  it  and then jumps onto the
left  branch  from  which  it is reinjected onto the branch of
large-amplitude  loops  and  the  cycle  closes.  As  $\gamma$
decreases  both  branches  fall  relative  to  the  bisectrix,
reflecting    the    gradual   growth   of   the   number   of
small-amplitude  loops  accompanied  by  a  reduction  in  the
number of large-amplitude loops.

In   Ref.~\cite{ring}   Ringland  {\it  et.  al.}  extensively
studied    the    properties    of    the   two-extremum   map
$x_{n+1}=Z(x_n)$ where
$$ Z(x)=(c+ax)(1-\tanh(sx))+(d+bx)(1+\tanh(sx)) .$$
At  small  values  of  the  parameter $s$ the map has a smooth
shape  similar  to  that  of  the cubic map. As $s$ increases,
the  slope  of  its  middle, negatively-inclined segment grows
steeper  while  the  $x$  distance  between  two map's extrema
vanishes.  As  $s$  tends  to  infinity,  the  Z-map  acquires
zigzag  shape  with  a  vertical middle segment. Ringland {\it
et.  al.}  have  shown that in the limiting case $s\rightarrow
\infty$  the  attractors  of the Z-map form Farey sequences as
other  map  parameters  are varied. They related this property
to  the  existence  of  a vertical segment since a decrease in
$s$  leads  to  a  gradual  transformation  of Farey sequences
into  U-sequences.  Although  the  dynamic origin of such maps
was  not  discussed,  a  possible relation of the Z-map to the
existence of mixed-mode oscillations was inferred.

One  can  easily  see  that  the  Z-map  in  its $s\rightarrow
\infty$  limit  is  qualitatively  similar to the map $f_Z(x)$
constructed  in  the  present  section.  In our map the middle
segment  is  always  vertical  due  to  the  presence  of  the
superunstable  orbit.  Applying  results  of Ringland {\it et.
al.}  to  system  (1),  we  may assume that the repulsive flow
defines  not  only  the  shape of mixed-mode orbits with their
nonoverlapping  bands  of  small-  and  large-amplitude loops,
but also their property to form Farey sequences.

\section{Conclusions}  \label{conc}

We  have  analyzed  in  detail  a  model exhibiting mixed-mode
oscillations.  By  constructing  the  model's slow manifold we
have  shown  that  mixed-modes  correspond  to periodic orbits
embedded  in  a  horseshoe-type strange set. This explains why
chaotic  oscillations  are  observed  in  transitions  between
adjacent   periodic  states.  The  organization  of  the  slow
manifold  into  a  horseshoe  also accounts for the signatures
of  chaos  in  transient  trajectories  when  the attractor is
periodic.

In  our  model  the mixed-mode periodic orbits do not lie on a
2-torus.  Following  transformations of the slow manifold from
its  simple  planar  organization  into  a  horseshoe, we have
also  shown  that  a  torus  does not exist as an intermediate
state  between  period-1  and  mixed-mode  oscillations. Thus,
our  scenario  for the emergence of mixed-mode oscillations is
an alternative to quasiperiodicity.

The   main  distinctive  feature  of  mixed-mode  states,  the
partition   of  their  orbits  into  nonoverlapping  bands  of
small-  and  large-amplitude  loops,  finds its explanation in
the  existence  of  the  repulsive  flow. Intersection of this
flow  with  the  slow  manifold  yields  a  complex  system of
superunstable  orbits.  The  position  of  the  slow  manifold
relative   to   the  repulsive  flow  determines  whether  the
formation  of  small  or  large  loops  of  the  attractor  is
favored.  As  parameters  change  along a path running through
the  mixed-mode  domain,  the  position  of the repulsive flow
changes  and  one  observes  sequences  of  mixed-mode  states
${\bf   n^1}\rightarrow  {\bf  1^1}\rightarrow  {\bf  1^n}  \;
({\bf  1^n}\rightarrow  {\bf  1^1}\rightarrow {\bf n^1})$ with
monotonously  increasing  (decreasing)  $S/L$  ratio.  As  the
results  of  Ringland  {\it  et.al.}  \cite{ring} suggest, the
fact  that  these  sequences are described by Farey arithmetic
is  also  a  consequence  of the presence of the superunstable
orbits.  Thus,  the  existence  of  the  repulsive flow places
this  horseshoe  into  a  separate  subclass  of  strange sets
whose  periodic  orbits  are given by $L^S$ mixed-mode states,
and  whose  windows  of  periodicity  are organized into Farey
sequences.

There  exist  indications  that  our  scenario may account for
the  mixed-mode  oscillations  observed  in  a number of model
and  experimental  studies.  This  appears  to  be  so for all
those  cases  in  which  the  mixed-mode  periodic states were
found  to  be  separated  by chaotic rather then quasiperiodic
oscillations.  As  Ringland  {\it  et.al.} \cite{ring} pointed
out,   zigzag-shaped  maps  with  a  vertical  middle  segment
indicative  of  a  superunstable trajectory were found in many
chemical and electrochemical systems.

\section{Acknowledgements}       
This  work  was  supported in part by a grant from the Natural
Sciences and Engineering Research Council of Canada.

\end{multicols}


\begin{thebibliography}{99}				      

\bibitem{ma-sw}  J.  Maselko and H. L. Swinney, J. Chem. Phys.
{\bf  85},  6430  (1986);  J. Maselko and H. L. Swinney, Phys.
Lett. A {\bf 119}, 403 (1987).
       
\bibitem{huds}  R.  A.  Schmitz,  K.  R.  Graziani,  and J. L.
Hudson,  J.  Chem.  Phys. {\bf 67}, 3040 (1977); J. L. Hudson,
M. Hart, and J. Marinko, J. Chem. Phys. {\bf 71}, 1601 (1979).
	      
\bibitem{gen}  V.  Petrov,  S.  K. Scott, and K. Showalter, J.
Chem.   Phys.   {\bf   97},  6191  (1992);  F.  Tracqui,  Acta
Biotheor.  {\bf  42},  147  (1994);  for  review on mixed-mode
oscillations   in   chemistry   see   I.  R.  Epstein  and  K.
Showalter, J. Phys. Chem. {\bf 100}, 13132 (1996).

\bibitem{roux}  P.  Richetti,  J.  C.  Roux, F. Argoul, and A.
Arneodo, J. Chem. Phys. {\bf 86}, 3339 (1987).

\bibitem{bark1}  D.  Barkley,  J.  Ringland, and J. S. Turner,
J. Chem. Phys. {\bf 87}, 3812 (1987).

\bibitem{lart}  C.  G. Steinmetz and R. Larter, J. Chem. Phys.
{\bf 94}, 1388 (1991).

\bibitem{albah}   F.   N.  Albahadily,  J.  Ringland,  and  M.
Schell, J. Chem. Phys. {\bf 90}, 813 (1989).

\bibitem{kop-gas}  M.  T.  M.  Koper  and P. Gaspard, J. Phys.
Chem.  {\bf  95},  4945 (1991); M. T. M. Koper and P. Gaspard,
J.  Chem.  Phys.  {\bf  96},  7797  (1992); M. T. M. Koper, P.
Gaspard,  and  J.  H.  Sluyters, J. Chem. Phys. {\bf 97}, 8250
(1992).

\bibitem{koper} M. T. M. Koper, Physica D {\bf 80}, 72 (1995).

\bibitem{ring}  J.  Ringland,  N.  Issa,  and M. Schell, Phys.
Rev. A {\bf 41}, 4223 (1990).

\bibitem{strik}  P.  E.  Strizhak  and  A.  L.  Kawczynski, J.
Phys. Chem. {\bf 99}, 10830 (1995).

\bibitem{foot1}  see  Ref.~\cite{koper}  for  another possible
extension of the Boissonade--De~Kepper model.

\bibitem{boiss}  J.  Boissonade  and P. J. De~Kepper, J. Phys.
Chem. {\bf 84}, 501 (1980).

\bibitem{foot2}  Polynomial  piece-wise  interpolation  of the
third   to  fifth  order  was  used  in  our  study.  For  the
particular   case   considered   in   this   section  accurate
approximation  for  the  entire  Poincar\'{e} section is given
by the polynomial $z(y)=.001y^3-.00432y^2-.22745y+.035977$.



\end{thebibliography}
\end{document}